\documentclass[sigconf,nonacm]{acmart}
\usepackage{multirow}
\usepackage{subfigure}
\usepackage[shortlabels]{enumitem}

\usepackage{balance}
\usepackage[ruled,linesnumbered]{algorithm2e}
\renewcommand\footnotetextcopyrightpermission[1]{} 
\AtBeginDocument{%
  \providecommand\BibTeX{{%
    \normalfont B\kern-0.5em{\scshape i\kern-0.25em b}\kern-0.8em\TeX}}}

\setcopyright{acmcopyright}
\copyrightyear{2022}
\acmYear{2022}
\acmDOI{XXXXXXX.XXXXXXX}

\hypersetup{draft}
\begin{document}

\title{MealRec: A Meal Recommendation Dataset}

\author{Ming Li}
\email{liming7677@whut.edu.cn}
\affiliation{
  \institution{Wuhan University of Technology}
  \city{Wuhan}
  \country{China}
}

\author{Lin Li}
\email{cathylilin@whut.edu.cn}
\affiliation{
  \institution{Wuhan University of Technology}
  \city{Wuhan}
  \country{China}
}

\author{Qing Xie}
\email{felixxq@whut.edu.cn}
\affiliation{
  \institution{Wuhan University of Technology}
  \city{Wuhan}
  \country{China}
}

\author{Jingling Yuan}
\email{yjl@whut.edu.cn}
\affiliation{
  \institution{Wuhan University of Technology}
  \city{Wuhan}
  \country{China}
}

\author{Xiaohui Tao}
\email{Xiaohui.Tao@usq.edu.au}
\affiliation{
  \institution{University of Southern Queensland}
  \city{Queensland}
  \country{Australia}
}



\begin{abstract}
Bundle recommendation systems aim to recommend a bundle of items for a user to consider as a whole. They have become a norm in modern life and have been applied to many real-world settings, such as product bundle recommendation, music playlist recommendation and travel package recommendation. However, compared to studies of bundle recommendation approaches in areas such as online shopping and digital music services, research on meal recommendations for restaurants in the hospitality industry has made limited progress, due largely to the lack of high-quality benchmark datasets. A publicly available dataset specialising in meal recommendation research for the research community is in urgent demand. In this paper, we introduce a meal recommendation dataset (MealRec) that aims to facilitate future research. MealRec is constructed from the user review records of Allrecipe.com, covering 1,500+ users, 7,200+ recipes and 3,800+ meals. Each recipe is described with rich information, such as ingredients, instructions, pictures, category and tags, etc; and each meal is three-course, consisting of an appetizer, a main dish and a dessert.  Furthermore, we propose a category-constrained meal recommendation model that is evaluated through comparative experiments with several state-of-the-art bundle recommendation methods on MealRec. Experimental results confirm the superiority of our model and demonstrate that MealRec is a promising testbed for meal recommendation related research. 
  
The MealRec dataset and the source code of our proposed model are available at https://github.com/WUT-IDEA/MealRec for access and reproducibility.
\end{abstract}

\ccsdesc[500]{Information systems~Recommendation systems}

\keywords{Datasets, Bundle, Meal Recommendation}

\maketitle

\begin{table*}[htb]
  \caption{Description of the data we have collected from Allrecipes.com and an example on ``Coconut Poke Cake''.}
  \label{tab:org_file}
  \small
\begin{tabular}{|c|l|l|l|}
\hline
\textbf{File name}                & \multicolumn{1}{c|}{\textbf{Field}} & \multicolumn{1}{c|}{\textbf{Description}}                                                                                    & \multicolumn{1}{c|}{\textbf{Example}}                                                                                                                                                                                                                                                              \\ \hline
\multirow{11}{*}{recipe.csv}      & recipe\_id                              & recipe identifier                                                                                                            & 7994                                                                                                                                                                                                                                                                                               \\ \cline{2-4} 
                                  & recipe\_name                            & the name of the recipe                                                                                                       & Coconut Poke Cake                                                                                                                                                                                                                                                                                  \\ \cline{2-4} 
                                  & aver\_rate                              & average user ratings(five-point scale)                                                                                       & 4.634286                                                                                                                                                                                                                                                                                           \\ \cline{2-4} 
                                  & image\_url                              & the url of the recipe image                                                                                                  & images.media-allrecipes.com/userphotos/720x405/334118.jpg                                                                                                                                                                                                                                          \\ \cline{2-4} 
                                  & category                                & the category the recipe belongs to in a meal                                                                                 & dessert                                                                                                                                                                                                                                                                                           \\ \cline{2-4} 
                                  & ingredients                             & ingredients included in the recipe                                                                                           & \begin{tabular}[c]{@{}l@{}}white cake mix; cream of coconut; sweetened condensed \\ milk; frozen whipped topping thawed; flaked coconut\end{tabular}                                                                                                                                               \\ \cline{2-4} 
                                  & cooking\_directions                     & food making process                                                                                                          & \begin{tabular}[c]{@{}l@{}}Prep 30 m Cook 1 h Ready In 2 h Prepare and bake white cake \\ mix according to package directions. Remove cake from oven...\end{tabular}                                                                                                                               \\ \cline{2-4} 
                                  & nutrition                               & nutritional content information of this food                                                                                 & \begin{tabular}[c]{@{}l@{}}sugars: \{hasCompleteData: True, name: Sugars, amount: 36.66, \\ percentDailyValue: 0, displayValue: 36.7, unit: g\}, …\end{tabular}                                                                                                                                    \\ \cline{2-4} 
                                  & review\_num                             & total number of the recipe’s reviews                                                                                         & 756                                                                                                                                                                                                                                                                                                \\ \cline{2-4} 
                                  & reviews                                 & all reviews of the recipe                                                                                                    & \begin{tabular}[c]{@{}l@{}}\{2765483: \{rating: 5, followersCount: 0, madeRecipesCount: \\ 535, dateLastModified: 2009-11-29T07:46:22.803, text: Using \\ the handle end of a wooden spoon perforates the cake better. \\ This cake is incredible for days!, followingCount: 0\}, ...\end{tabular} \\ \cline{2-4} 
                                  & tags                                    & \begin{tabular}[c]{@{}l@{}}several tags selected by the creator \\ represent features of the recipe\end{tabular} & \begin{tabular}[c]{@{}l@{}}60-minutes-or-less; desserts;eggs-dairy; fruit; oven; potluck; \\ picnic; cakes; nuts; eggs; dietary; coconut; to-go; equipment\end{tabular}                                                                                                                            \\ \hline
\multirow{4}{*}{user\_recipe.csv} & user\_id                                & user identifier                                                                                                              & 168192                                                                                                                                                                                                                                                                                             \\ \cline{2-4} 
                                  & recipe\_id                              & recipe identifier                                                                                                            & 7994                                                                                                                                                                                                                                                                                               \\ \cline{2-4} 
                                  & rating                                  & user ratings for the recipe(five-point scale)                                                                                & 5                                                                                                                                                                                                                                                                                                  \\ \cline{2-4} 
                                  & dateLastModified                        & the time the user review was last modified                                                                                   & 2014-04-25T14:54:20                                                                                                                                                                                                                                                                                \\ \hline
\end{tabular}
\end{table*}

\section{Introduction}
Compared to traditional recommendation systems that make single item recommendations, the bundle recommendation aims to recommend a bundle of items for a user to consider as a package. Bundle recommendation systems predict users' preferences for a bundle of items rather than individual items~\cite{bundle_first}, which has become an important recommendation technology in modern life. 
Products provided by sellers are increasing each day in both quantity and variety, meaning consumers are living in a world with overwhelming information~\cite{rec_survey_new}. Bundle recommendation systems can help users quickly identify a high-quality set of items that meet their needs and interest, and a result, these systems help alleviate burdens on users caused by information overload. Additionally, bundle recommendation systems can increase the visibility of low-exposure items in bundles, thus reducing the long-tail effect of products and increase sales for sellers~\cite{long_tail}. 

Because of the great practical and commercial potential, more and more research is being conducted on bundle recommendations, with significant progress on many real-world settings, such as product bundle recommendations \cite{bundle_first,product_bundle_1,product_bundle_2,product_bundle_3,product_bundle_4,bundle_steam,RGNN}, music playlist recommendations \cite{Netease_dataset,BGCN,music_playlist_1,music_playlist_2,music_playlist_3}, and travel package recommendations \cite{tour_dataset,travel_1,travel_2}, etc. Personalized bundle recommendation technology can help users quickly locate high-quality bundles matching their interest, improve user experience, and benefit both consumers and sellers. The industry has also realized the great potential of bundle recommendation systems, evidenced by their applications in online shopping (e.g., Amazon), music platforms (e.g., Spotify, NetEase music), and travel services (e.g., Ctrip), along with many other areas. 

The research on meal recommendation, however, has made little progress compared to other well-studied bundle recommendation problems. Studies have revealed that the research community lacks a quality benchmark dataset specializing in meals for the study of recommender systems. A high-quality benchmark dataset significantly facilitates and stimulates research in an area. This phenomenon is observed in Amazon product data\footnote{http://jmcauley.ucsd.edu/data/amazon/} for product bundle recommendation, the Spotify dataset and Netease Cloud Music dataset~\cite{Netease_dataset,music_playlist_3} for music playlist recommendation, travel route dataset~\cite{tour_dataset} for travel package recommendation and Recipe 1M data for food oriented cross-modal retrieval~\cite{cikm1,cikm2}. Because of these datasets, each respective field has benefited significant from the research work conducted using the high-quality data. 
Meal recommendation is an application of bundle recommendation in food scenarios. Its task is to recommend recipes that are enjoyed as a meal for a user. However, there are few existing studies on meal recommendations, and most of them are conducted on proprietary datasets \cite{meal_1,meal_3,meal_6} or nutrition datasets \cite{meal_2,meal_4}. These datasets are either not publicly available for access or lack user-meal interaction data to empower the training of data-intensive models. Compared with the other bundle recommendations previously mentioned, the lack of public datasets in this space has resulted in the lag of meal recommendation. Therefore, a public meal recommendation dataset is in urgent demand for research in this field.

Responding to this demand, we introduce a dataset named MealRec for meal recommendation in this paper. We collected recipes and the related user reviews from Allrecipes.com \footnote{https://www.allrecipes.com}, one of the most popular food sharing sites in the world. Recipe is an important representation of food, which contains rich descriptions such as food name, ingredients, instructions, pictures, etc. We propose a data construction method to construct meals and user-meal interactions. Each of the meals consists of an appetizer, a main dish and a dessert in the form of three-courses \footnote{https://en.wikipedia.org/wiki/Full-course\_dinner}\,\,\footnote{https://www.collinsdictionary.com/us/dictionary/english/three-course-meal}. The meal construction is based on the structural constraints of the recipe categories, and considers the respective consistency of the explicit and implicit features of a meal. Finally, the MealRec dataset contains 1,500+ users, 7,200+ recipes and 3,800+ meals. Furthermore, we propose a category-constrained meal recommendation model and conduct comparative experiments using MealRec with several state-of-the-art meal and general bundle recommendation methods including HR@K and NDCG@K. The results prove the superiority of our model and demonstrate MealRec as a good testbed for meal recommendation related research. The dataset and implementation of the proposed model have been made available to the public~\footnote{https://github.com/WUT-IDEA/MealRec}, aiming to facilitate further research on meal recommendation.

The contributions delivered by this work are summarized below:
\begin{itemize}
  \item A new dataset is introduced for meal recommendation research, which is of quality and the first one with specialty in meals. The dataset is made publicly available for free access by the research community.   
  \item An innovative category-constrained bundle recommendation model is proposed with promising performance proven through an empirical experiment.
  \item A new paradigm is provided for evaluating bundle recommendation methods, especially those on meals, with the new MealRec dataset, which will guide other related research in the future.
\end{itemize}

The left of this paper is organized as follows. Section~\ref{sec-RW} focuses on related research on bundle recommendation including meal recommendation and related datasets. In Section~\ref{sec-dataset}, we explain the construction of MealRec with related analysis. In Section~\ref{sec-model}, we propose a category-constrained meal recommendation model against the meal recommendation problem. Section~\ref{sec-exp} will report the results of comparative experiments on MealRec. Finally, Section~\ref{sec-conclusion} makes conclusion to the paper.

\section{Related Work}\label{sec-RW}
\subsection{Bundle Recommendation}
Bundle Recommendation problem refers to predict a user’s preference on a bundle rather than an individual item~\cite{bundle_first}. Bundle recommendations with great practical and commercial value have been applied to some real-world services, such as product bundle recommendation, music playlist recommendation and travel package recommendation, etc. Among bundle recommendations, product bundle recommendations are the most widely studied. Association analysis techniques \cite{product_bundle_3} mine the correlation between different projects, providing interesting product bundles such as diapers and beer. Traditional recommendation techniques like CF are also applied to this task. Liu et al.~\cite{product_bundle_2} estimated the possibility that consumers would purchase recommended items together with items that have already been purchased.
Many researchers~\cite{BGCN,package_rec,RGNN,MIDGN} have also made great progress on the application of graph convolution technology in product bundle recommendation. BGCN~\cite{BGCN} utilizes the graph neural network’s power to learn the representation of user and bundle from complex structures. 

Music playlist recommendations \cite{Netease_dataset,BGCN,music_playlist_1,music_playlist_2,music_playlist_3} recommend albums and user generated lists to listeners. Deep learning methods are often used to aggregate the song features to obtain the representation of the playlist. AGREE \cite{AGREE} uses NCF \cite{NCF} to solve the bundle recommendation problem from the perspective of neural representation learning. DAM \cite{DAM} jointly modeled user-item interactions and user-bundle interactions in a multi-task learning model. AttList \cite{AttList} leverages the hierarchical structure of items, lists, and users to capture the containment relationship between lists and items, revealing additional insights into user preferences. 

Travel package recommendations is a special application in bundle recommendations. They select points of interest (POI) to visit in unfamiliar cities by selecting POIs that align with user interest preferences and trip constraints. The PersTour algorithm \cite{tour_dataset} considers both POI popularity and user interest preferences to recommend suitable POIs to visit and the amount of time to spend at each POI. In addition, Herzog et al. \cite{travel_2} list attributes that influence people’s decisions on the choice of walking routes, such as air quality, road damages and lack of accessibility. This eco-sensitive smart tourism application provides support for tourists. 

The development of bundle recommendation research is greatly dependent upon the support of public datasets. Amazon product data
for product bundle recommendation and travel route dataset\cite{tour_dataset} for travel package recommendation have greatly promoted the development of related research. 
The music playlist datasets NetEase \cite{Netease_dataset} and Spotify \cite{music_playlist_3}, and the booklist datasets Youshu \cite{DAM} and Goodreads \cite{music_playlist_3}, are also often used in experiments to compare the performance of bundle recommendation models. These datasets all provide convenience for bundle recommendation research. However, none of these datasets have the feature of category constraints, which is very important for a meal recommendation dataset. For example, a user cannot have three starters or drinks as his meal. Therefore, category information is important and indispensable for the meal recommendation task, which is also not available in these datasets.

\subsection{Meal Recommendation and Related Datasets}
Meal recommendations been subject to research attention for a long time. Elsweiler \cite{meal_1,meal_1_1} attempted to incorporate health and nutrition into the meal recommendation problem for assisting people achieve a healthy diet. DIETOS (DIET Organizer System) \cite{meal_2} has provided personalized dietary recommendations by analyzing the consumption data of healthy people and diet-related chronic disease patients. Multi-objective optimization technology has been introduced into the RS to do healthy meal recommendation by considering healthy nutrients, harmonization, and coverage of ingredients in the pantry \cite{meal_3}. Depending on different situations, its goal can be achieved by adjusting the optimization objective. In addition, evolutionary approach is also introduced into the meal recommendation \cite{meal_4}. It recommends diet plans or training packages to users based on user preferences and goals, aiming to provide users with a more comprehensive experience. The aforementioned methods mainly make recommendations to users for health or specific nutritional needs, and pay less attention to users' personalized preferences. It is very difficult to promote and apply a recommendation system that does not cater to users’ food preferences. 
GFR is \cite{meal_5} a general collaborative learning framework of personalized recommendation, which introduces rich contextual information. However this method does not consider the affiliation information between the dishes and the meal, which is very important for the representation of meals and the learning of user preferences.

Compared to other fields such as movies and e-commerce, it is difficult to collect large-scale user-meal interaction data. This is mainly due to the lack of a unified ordering platform, which means large-scale user-meal interaction data cannot be directly obtained. However, the data obtained from a single restaurant is sparse, and is insufficient to support the research data-intensive models. 
To collect food preference data of users, Elsweiler \cite{meal_1,meal_1_1} created a food portal website. 
The growth of internet information has also given some works \cite{meal_3,meal_6} the opportunity to obtain food data from food social networking sites such as Allrecipes.com. However none of these works have published experimental datasets. 
The researches \cite{meal_2,meal_4} use officially released nutritional databases such as the one\footnote{www.healthcanada.gc.ca/cnf} released by the minister of health Canada, but these datasets do not have the interaction data that is important in recommendation. 
Unfortunately, literature suggests there is currently no public meal recommendation dataset which contains user-meal interaction data. 
The absence of public benchmark datasets has largely caused the research on meal recommendation to lag behind other bundle recommendation studies.

To facilitate research on meal recommendation, we construct an original meal recommendation dataset (MealRec) and make it open source. In the next section, we elucidate its construction process and perform data analysis on it.

\section{{MealRec}: A New Dataset for Meal Recommendation}\label{sec-dataset}

MealRec builds meal data and user-meal interaction data based on recipe data and user-recipe interaction data. 
In this section, we first explain the recipe collection, and then propose a method for constructing reasonable meals and user-meal interactions, and finally analyze the dataset. 

\begin{figure*}[!h]
  \centering
  \includegraphics[width=11cm]{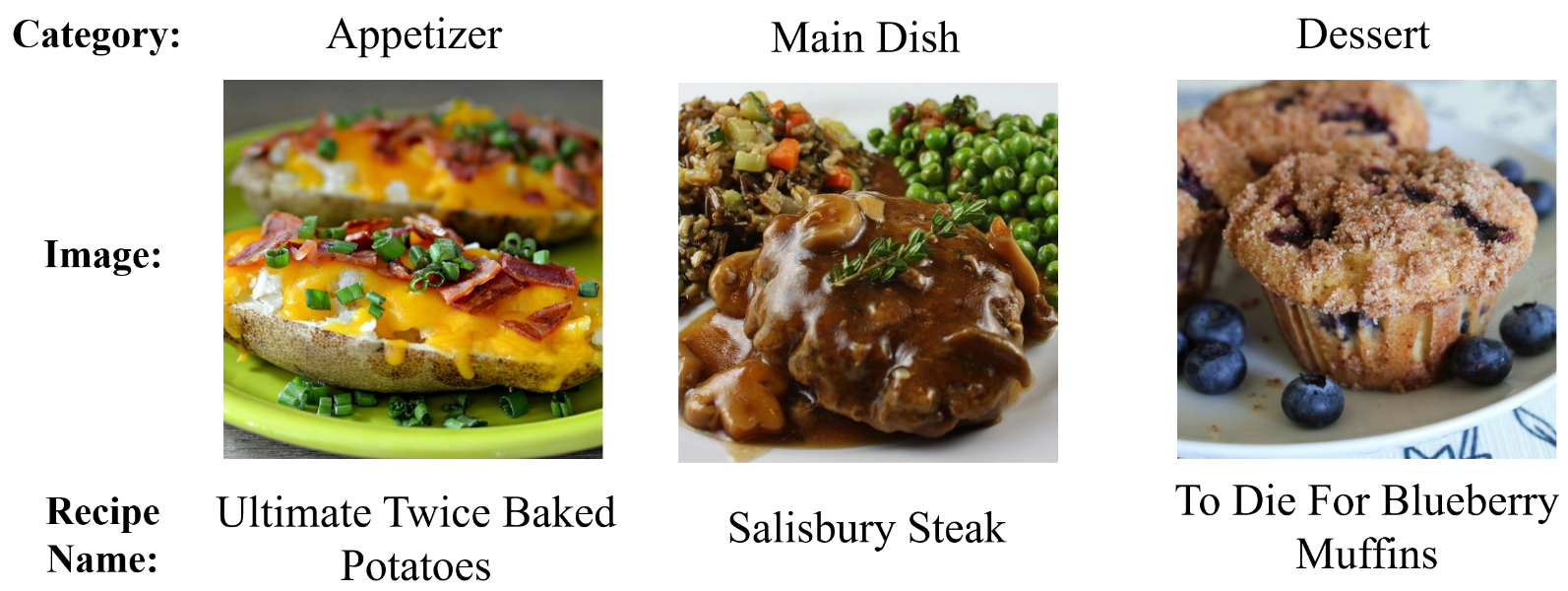}
  \caption{An example of constructed meal.}
  \label{example:meal}
\end{figure*}

\subsection{Recipe Collection}
 Because recipes are an important representation of food and can be easily obtained from recipe sharing platforms such as Allrecipes.com and Food.com, recipe data is chosen to construct our dataset. We get recipe data and corresponding user reviews from Allrecipes.com\footnote{https://www.allrecipes.com} between 2000 and 2018. 
 If the user's rating for the recipe is greater than or equal to 4 (on a five-point scale), we judge that the user likes the food, otherwise, a lower rating means we consider that the user does not like it (including user unseen recipes). In this paper, we only consider positive feedback from users and filter out user reviews with ratings less than 4.
 In order to clear the invalid data that is too sparse, each user is required to interact positively with at least 5 recipes and each recipe has been commented on at least 5 times.
 We obtained 30,833 recipes and 3,759,181 user-recipe interaction records after basic data processing and filtering. 
 
 Table \ref{tab:org_file} shows the structure of the original data and gives an example of "Coconut Poke Cake". 
 As shown in the table, the recipe data contains rich descriptive information, such as ingredients, instructions, pictures, category and tags, etc. In particular, the category field is determined by us based on the order in which the food is served in a meal, such as main dish, dessert, appetizer, etc. And the tag field is attached by the creator of the recipe, indicating features of the food. The interaction data includes the rating and timestamps of each user's reviews on recipes, from which we can obtain a timeline of the user's recipe interactions. After completing this data collection, we start the construction of MealRec.

\subsection{Meal Construction}

Many cultures enjoy full-course meals with a unique assortment of food and customs. Full-course meals have a rich, diverse history from many regions that is a result of evolving food trends over time. Many meals only contain one course, but meals can feature up to 12 or more courses. The basic full-course meal\footnote{https://en.wikipedia.org/wiki/Full-course\_dinner}\,\,\footnote{https://www.collinsdictionary.com/us/dictionary/english/three-course-meal} is made up of three courses: an appetizer, a main dish, and a dessert. Also known as a three-course meal or a standard course meal, you will sometimes see restaurants offering a full menu with these three items. In this paper, we construct the meal with reference to the composition of the basic and regular three-course meal, i.e. a meal that consists of an appetizer, a main dish, and a dessert. It's worth mentioning that other forms of meal can be constructed in the same way. After clarifying the composition of the meal, we first filter the recipes and their related interaction data that do not belong to these three categories in the original data, and then try to construct meals.

A meal course is a set of food items served at once, following a theme, flavor, requirement or otherwise. Therefore, we need to consider that all courses in a meal have similar features when constructing meals.
On the user side, each user has his own specific preferences and favorite tastes, and his favorite foods share implicit features. For example, for a user who likes to eat fried chicken and chocolate cake, the food he or she likes implicitly contains a high calorie feature. 
Especially over a period of time, a user's favorite foods may have more significant commonalities, because some foods only appear in a specific season and a user's preferences in a short period of time will be more consistent.
On the creator side, the tags of a recipe are selected by the creator when uploading the recipe and they represent explicit features of the food. 

In order to construct reasonable meals, we consider both the implicit and explicit features of the meal to be consistent. Lines 3 to 20 of the Algorithm \ref{algo:construct} illustrate this process. 
Because the all recipes a user interacts with over a period of time follow the same implicit features. Every time we construct a meal, the recipes in the meal is required to be enjoyed by a user over a period of time, which guarantees the consistency in implicit features of the meal. 
Then, The number of shared tags of the three recipes is required to be greater than 0 to ensure the consistency of the explicit features of the meal. Figure \ref{example:meal} shows an example of a constructed meal that contains recipes that share the tags 'North-American', 'American', and 'Easy'. 

After constructing meals, we start constructing user-meal interactions. Its construction process is shown in lines 21 to 30 of the algorithm \ref{algo:construct}. Bundle recommendations in some scenarios are given high user tolerance, such as a music playlist recommendation. A user who likes most of the songs in a playlist is likely to like the playlist, even if he or she doesn't like a few songs in it. But users have lower tolerance for a meal recommendation and a user won't choose a meal with a food they don't like. Therefore, in the process of building user-meal interactions, we only consider that the user likes a meal when the user likes all the courses in the meal. 
After the output of the algorithm , we obtain the original constructed meal and user-meal interaction. 
We further processed our dataset by retaining meals appearing in at least in 5 user-meal interactions and users consuming at least 5 meals. As shown in Table \ref{tab:meal}, after completing the build, we produced two new data files. 

\begin{algorithm}[htb]
\footnotesize
 \caption{A construction method for meal and user-meal interaction}
 \label{algo:construct}
 \LinesNumbered
 \KwIn{Recipe $R$, User-Recipe $UR$}
 \KwOut{Meal $M$, User-Meal $UM$}
    $UM \gets  M \gets\left \{  \right \} $\;
    $U  \gets Unique(UR.u) $\;
 \tcc{meal construction}
 $i \gets 0 $\;
 \For{ u \  \textbf{in} \ U }{
     $R_u \gets GetInteractedRecipesOfUser\left ( UR, u \right )$\;

  \While{ $AppetizerNum\left ( R_u \right )\ne 0 $ $ \textbf{or} \ MainDishNum\left ( R_u \right )\ne 0 $ $  \textbf{or} \ DessertNum\left ( R_u \right )\ne 0  $} {
   $  r_{1} \gets GetRandomAppetizer \left ( R_u \right )$\;
   $  r_{2} \gets GetRandomMainDish \left ( R_u \right )$\;
   $  r_{3} \gets GetRandomDessert \left ( R_u \right )$\;
   $  t_{1} \gets GetInteractionTimestamp \left ( UR, u, r_{1} \right )$\;
   $  t_{2} \gets GetInteractionTimestamp \left ( UR, u, r_{2} \right )$\;
   $  t_{3} \gets GetInteractionTimestamp \left ( UR, u, r_{3} \right )$\;
   \tcp{require three recipes to be enjoyed by a user over a period of time}
   \If{$Max\left ( t_{1}, t_{2}, t_{3} \right ) - Min\left ( t_{1}, t_{2}, t_{3} \right ) < \lambda $}{
   \tcp{require at least one tag to be shared}
    \If{ $GetNumOfSharedTag(r_{1}, r_{2}, r_{3}) \ge 1$ }{
	  $M \left [ i++ \right ]  \gets \left \{ r_{1}, r_{2}, r_{3} \right \}$\;
      $R_u.remove\left (r_{1}, r_{2}, r_{3} \right )$\;
    }
   }
  }
 }
 \tcc{user-meal interaction construction}
 $i \gets 0 $\;
 \For{ u \  \textbf{in} \ U  \textbf{and} m \  \textbf{in} \ M}{ 
  $R_u \gets GetInteractedRecipesOfUser\left ( UR, u \right )$\;
  \For{ \textbf{and} m \  \textbf{in} \ M}{
   \tcp{require all recipes in meal to be liked by users}
   \If{$ m.recipes \in  R_u  $ }{
    $UM\left [ i++ \right ] \gets \left \{u,m \right \}$\;
    $R_u.remove\left (m.recipes \right )$\;
   }
  }
 }
  return  M, UM

\end{algorithm}

\begin{table}[htb]
 \caption{Description of constructed meal and user-meal interaction data.}
  \label{tab:meal}
  \small
\begin{tabular}{|c|l|l|}
\hline
\textbf{File Name}              & \multicolumn{1}{c|}{\textbf{Field}} & \multicolumn{1}{c|}{\textbf{Description}} \\ \hline
\multirow{4}{*}{meal.csv}       & meal\_id                                & meal identifier                           \\ \cline{2-3} 
                                & appetizer\_id                           & recipe identifier of this appetizer       \\ \cline{2-3} 
                                & main\_dish\_id                          & recipe identifier of this main dish       \\ \cline{2-3} 
                                & dessert\_id                             & recipe identifier of this dessert         \\ \hline
\multirow{2}{*}{user\_meal.csv} & user\_id                                & user identifier                           \\ \cline{2-3} 
                                & meal\_id                                & meal identifier                           \\ \hline
\end{tabular}
\end{table}

\begin{table}[htb]
 \caption{Dataset statistics of MealRec and Youshu.}
  \label{tab:statistics}
  \small
\begin{tabular}{|l|l|l|}
\hline
\textbf{Dataset}                  & \textbf{MealRec} & \textbf{Youshu}   \\ \hline
\# user                           & 1,575             & 8,039            \\ \hline
\# bundle                  & 3,817             & 4,771            \\ \hline
\# item                  & 7,280             & 32,770           \\ \hline
\# user-bundle              & 46,767            & 51,377           \\ \hline
\# user-item              & 151,148           & 138,515          \\ \hline
\# bundle-item      & 11,451            & 176,667          \\ \hline
\# avg. bundle interactions for user & 29.69             & 6.39             \\ \hline
\# avg. item interactions for user & 95.96             & 17.23            \\ \hline
\# avg. bundle size         & 3                 & 37.03            \\ \hline
\# user-item density      & 1.30\%            & 0.05\%           \\ \hline
\# user-bundle density      & 0.77\%            & 0.13\%           \\ \hline
\end{tabular}
\end{table}

\subsection{Data Analysis}

The detailed statistics of MealRec are shown in Table \ref{tab:statistics}. Based on the previously mentioned construction method, this dataset contains 1,575 users, 3,817 meals, 7,280 recipes, 46,767 user-meal interactions, and 151,148 user-recipe interactions. In Table \ref{tab:statistics}, we compare MealRec with a real-world dataset Youshu \cite{DAM}, which is a book list dataset. The scales of the two dataset are in the same order of magnitude. The data density of MealRec dataset is higher than that of YouShu. One reason is the size of the bundles, as smaller bundles are more likely to interact than larger ones. Another reason may be that the strict data construction conditions cause relatively sparse data to be filtered. In general, MealRec has comparable data size and higher data density to the real-world dataset, and it can empower the training of data-intensive meal recommendation models. 

\begin{figure}[htb]
\centering
\subfigure[Number of recipe interactions per user]{
\includegraphics[width=3.8cm]{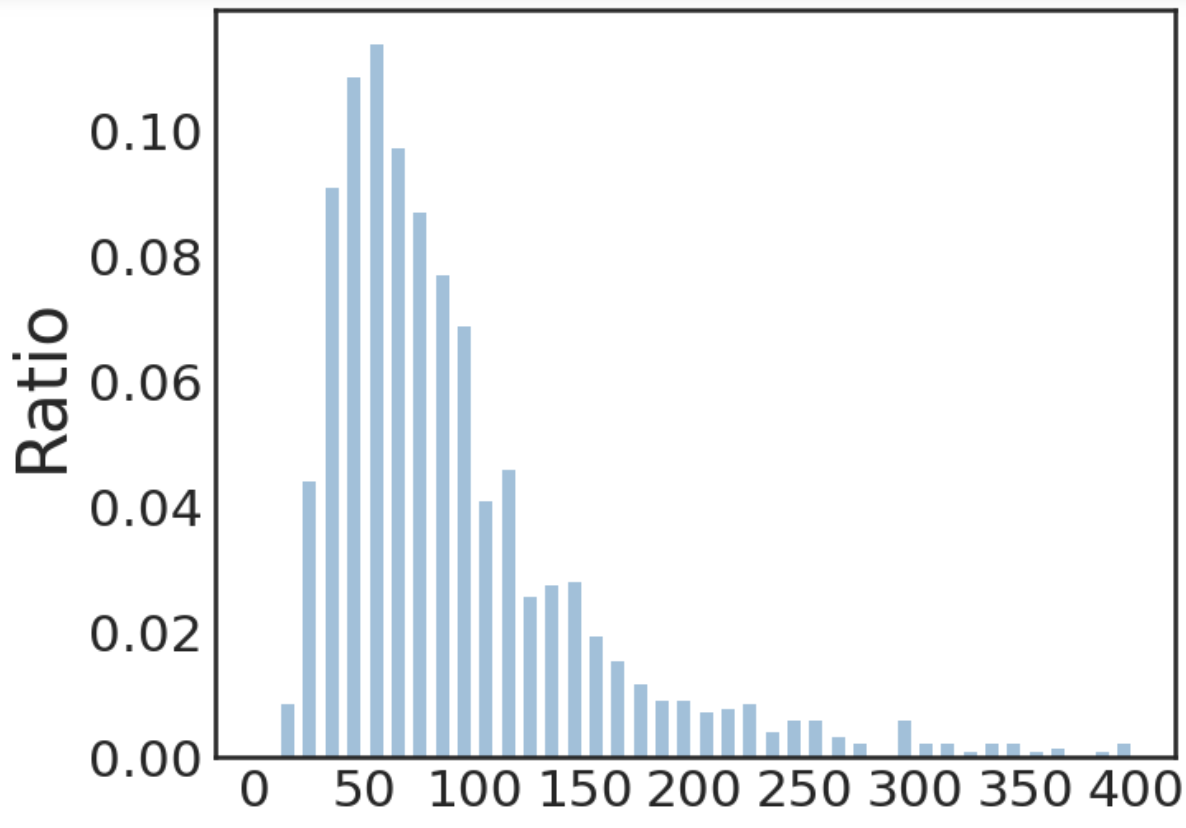}
}
\quad
\subfigure[Number of user interactions per recipe]{
\includegraphics[width=3.8cm]{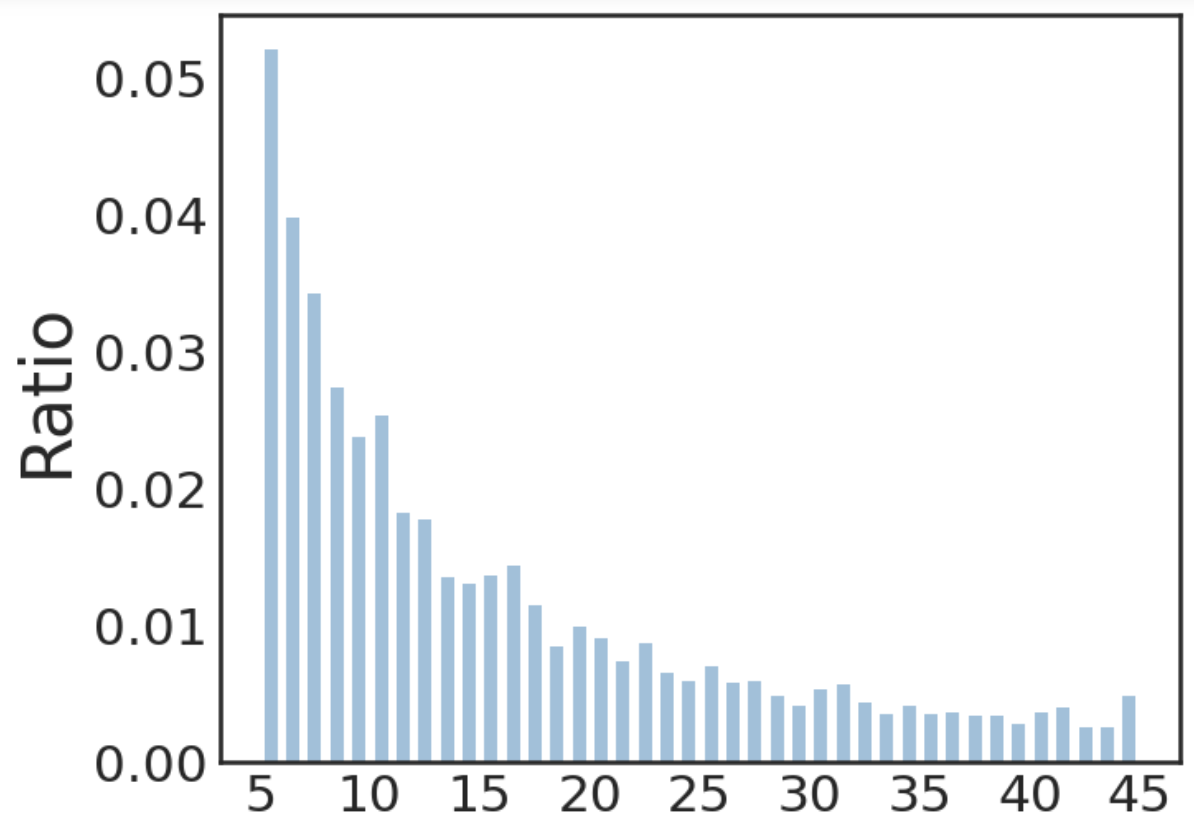}
}
\quad
\subfigure[Number of meal interactions per user]{
\includegraphics[width=3.8cm]{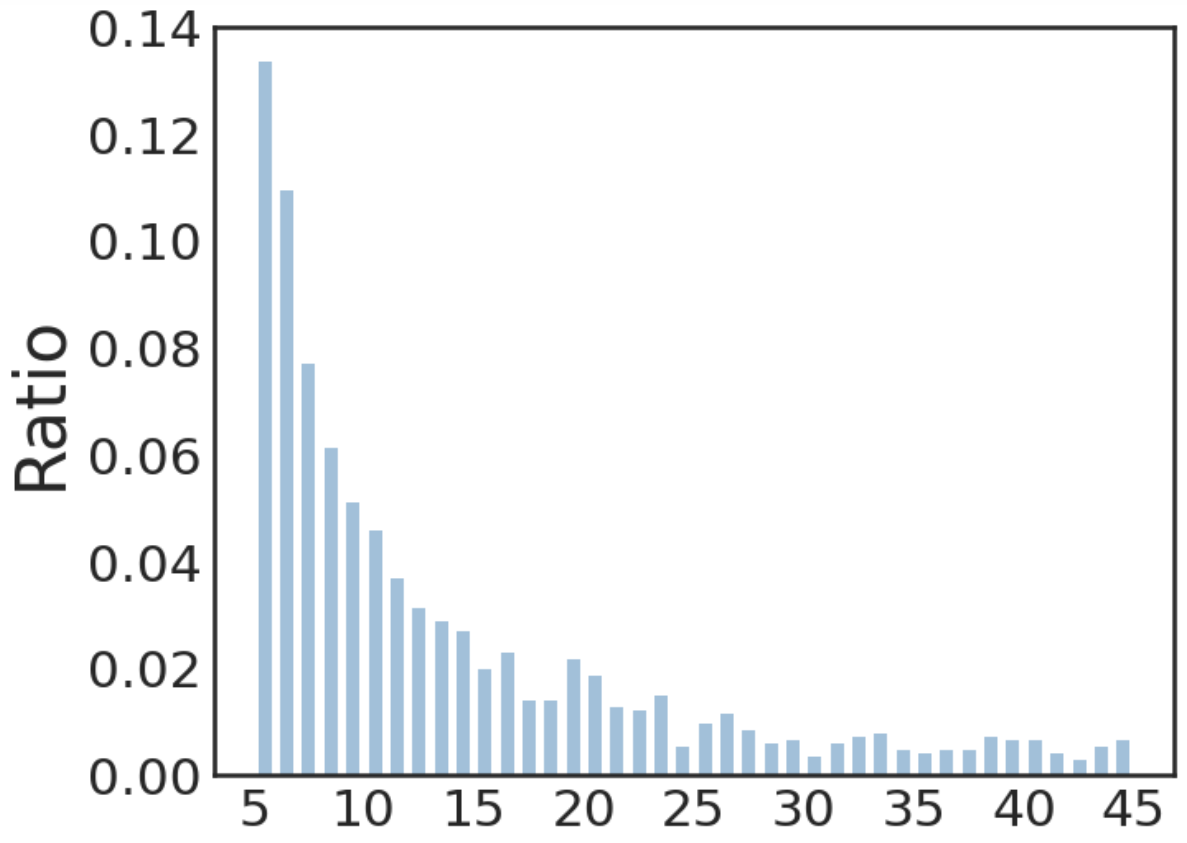}
}
\quad
\subfigure[Number of user interactions per meal]{
\includegraphics[width=3.8cm]{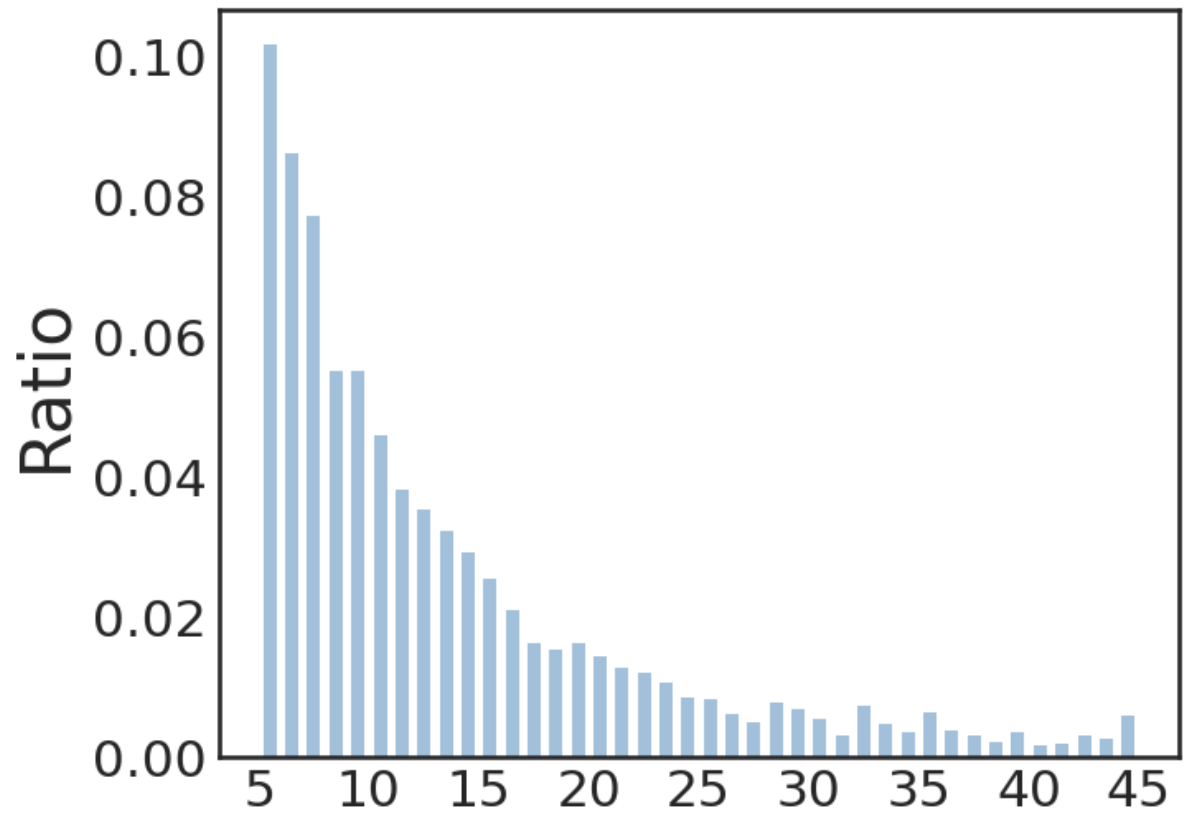}
}
\caption{Data distributions of MealRec.}
\label{distributions}
\end{figure}

 Figure \ref{distributions} shows the data distribution of MealRec. Note that we discarded users who interacted with less than 5 meals or recipes, and discarded meals and recipes that were interacted with less than 5 times when creating MealRec. We assume that data is too sparse to be informative, or just noise. 
 The data distribution of the number of user interaction recipes is shown in Figure \ref{distributions}(a). The majority of users are roughly distributed in the interval 40 to 80. This is because users with fewer recipe interactions did not construct enough user-meal interactions and were discarded. 
 Among the 7,280 recipes included in MealRec, there are 2,737 appetizers, 2,552 main dishes and 1,991 desserts. 

\section{Category-Constrained Meal Recommendation: A New Model}\label{sec-model}

\subsection{Problem Definition}

The meal recommendation problem can be formulated as follows. 
We denote the set of users, recipes and meals as $\mathcal{U}$ and $\mathcal{R}$ and $\mathcal{M}$ where the size of these sets is $|\mathcal{U}|$, $|\mathcal{R}|$ and $|\mathcal{M}|$. Every meal in $\mathcal{M}$ is composed of recipes from $\mathcal{R}$, and the affiliation relationship between $\mathcal{R}$ and $\mathcal{M}$ is denoted as recipe ID  matrix $\boldsymbol{A} \in \mathbb{N}^{|\mathcal{R}| \times |\mathcal{C}|}$, where $\mathcal{C}$ is the category set of recipes and $|\mathcal{C}|$ is its size. Each meal consists of an appetizer, a main dish, and a dessert, thus $\mathcal{C} = \left \{ appetizer,\ main\ dish,\ dessert \right \}$(abbreviated as $\left \{ a,\ s,\ d \right \}$). 
The corresponding matrix of recipes and categories can be defined as $\boldsymbol{G} \in \left \{ 0, 1 \right \} ^{|\mathcal{R}|\times |\mathcal{C}|}$, where $g_{rc} = 1$ indicates recipe $r$ belongs to category $m$, otherwise it does not.
We define the user-meal interaction matrix as $\boldsymbol{T} \in \left \{ 0, 1 \right \} ^{|\mathcal{U}|\times |\mathcal{M}|}$, where $t_{um}$ indicates the feedback from $u$ to $m$. We let $t_{um} = 1$ indicate $m$ has positive feedback from $u$ and $t_{um} = 0$ otherwise.

The meal recommendation problem can be defined as input from the users $\mathcal{U}$, recipes $\mathcal{R}$, meals $\mathcal{M}$, the recipe-meal affiliation matrix $\boldsymbol{A}$, recipe-category corresponding matrix $\boldsymbol{G}$, and the user-meal interaction matrix $\boldsymbol{T}$. Given a user and a meal, estimate preference of $u$ to $m$, denoted as $\hat{r}_{um}$.

\subsection{Research Design}

Existing methods \cite{AttList,BGCN,MIDGN} usually learn user preferences by propagating bottom-up from the item level to the bundle level, and finally to the user level.
However, if these methods are applied to the meal recommendation task, they suffer from two drawbacks. First, these methods heavily focus on meal-wise user preferences, ignoring important category information. Users may have specific preferences for specific categories, such as a user might prefer a fried one for appetizer and a cold one for desserts. 
Second, these methods also fail to take into account the different importance of the recipes aggregated to predict user preferences for target meals. Since including irrelevant information can introduce noise and may harm the final user representation quality, it is necessary to reduce irrelevant information in user preferences and focus on the information that pertains most to the target meal. Keeping this concern in mind, we have three observations for the meal recommendation task on MealRec:
\begin{itemize}
  \item \textbf{Meal-wise user preferences.} Users and meals(recipe bundles) naturally form a hierarchical structure that could be helpful for modeling user preferences in meal-wise. 
  \item \textbf{Category-wise user preferences.} Each meal is category-constrained, and each recipe belongs to a specific category. The hierarchical structure between recipes, categories and users can be used to learn category-wise user preferences that additionally contains category information compared to meal-wise user preferences.
  \item \textbf{Different recipes aggregated into user preferences have different importance in predicting the user's preference for the target meal.} When meal-wise and category-wise user preferences, which aggregate rich information from recipes, are used for recommendation, not all information in the user preference is of equal use to judging whether the target user likes the target meal. Only those target meal related information provides us a ground for analysis and indication to the preference. Thus, different recipes should be treated with different weights taking into consideration of the target meal when we aggregate recipes to obtain user preferences. 
\end{itemize}

\begin{figure}[t]
  \centering
  \includegraphics[width=\columnwidth]{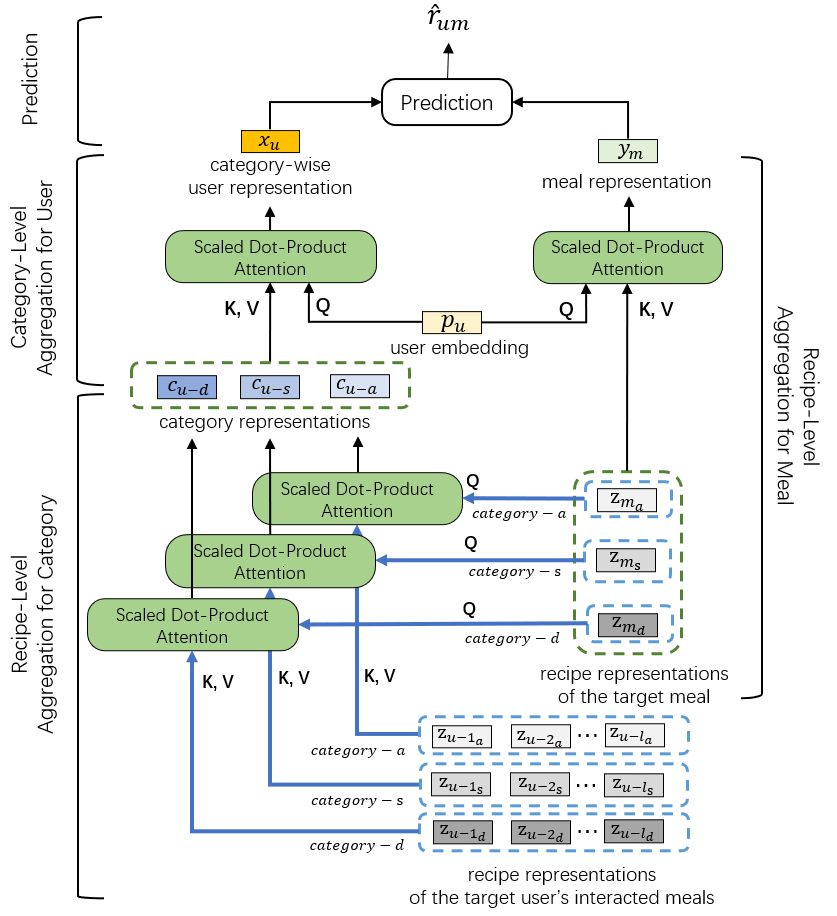}
  \caption{The framework of our proposed model.}
  \label{framework}
\end{figure}

Based on the observations, we propose an innovative category-constrained meal recommendation model (\textbf{CCMR}). First, our model follows the hierarchical structure of recipes, categories, and users to learn category-wise user preferences, revealing additional category insights into user preferences. Second, to reduce irrelevant information and focus on the most relevant information and target meal, the attention mechanism is used to extract the information from the user's favorite recipes based on the target meal and aggregates them by category. Then, we use the attention mechanism to perform category-level aggregation to obtain category-wise user preferences. 
Figure \ref{framework} illustrates our proposed model, which is made up of the meal representation learning, the category-wise user preference representation learning, and the prediction.

The goal of the our model is to learn latent representations for the target user $x_u$ and the target meal $y_m$, and then use them to predict user preference scores $\hat{r}_{um}$ on the target meal. 

\subsection{Meal Representation}

A meal is composed of multiple recipes, and these recipes together reflect the meal. We obtain representation of the target meal by aggregating representations of the recipes in it, as shown in the right half of Figure \ref{framework}. Next, we explain its details.

\subsubsection{Recipe Representation}
The recipe representation (omitted in Figure \ref{framework}) aims to transform each recipe in the target meal from a recipe ID to a continuous vector embedding. 
We first create a learnable recipe embedding matrix $E \in \mathbb{R}^{|\mathcal{R}|\times d}$ where $d$ is the latent dimensionality. Based on $E$, the recipe with category $t \in \mathcal{C}$ in the target meal $m$ can retrieve its embedding $e_{m_t} \in \mathbb{R}^{1\times d}$. 
Considering that different categories of recipes have their unique features, we also create a learnable category embedding matrix $O \in \mathbb{R}^{|\mathcal{C}|\times d}$ and $o_{m_t} \in \mathbb{R}^{1\times d}$ denotes the category representation for $e_{m_t}$. We further add them together as the final recipe representation:

\begin{equation}
\setlength{\abovedisplayskip}{3pt}
\setlength{\belowdisplayskip}{3pt}
\label{zmc}
z_{m_t} = e_{m_t} + o_{m_t}
\end{equation}

After embedding, the three recipes of the meal can be represented as Eq(\ref{zm}). Because each category in each meal contains only one recipe, $z_m$ is also the category representations of the target meal.

\begin{equation}
\setlength{\abovedisplayskip}{3pt}
\setlength{\belowdisplayskip}{3pt}
\label{zm}
z_m = 
\begin{bmatrix}
 z_{m_{a}}\\
 z_{m_{s}}\\
 z_{m_{d}}
\end{bmatrix} \in \mathbb{R}^{|\mathcal{C}|\times d}
\end{equation}

Similarly, the target user $u$ can be mapped by user ID to a user embedding $p_u \in \mathbb{R}^{1\times d}$ through a learnable embedding matrix $P \in \mathbb{R}^{|\mathcal{U}|\times d}$. 

\subsubsection{Recipe-Level Aggregation for Meal}
After getting the representation of the recipes in the target meal, we aggregate them to obtain the representation of the target meal. Considering that each user has different emphasis on the recipes when selecting meals, the scaled dot-product attention \cite{dot_product_att} is used to aggregate recipe representations to obtain personalized meal representations. The definition of the scaled dot-product attention is Eq(\ref{att_Eq}):

\begin{equation}
\setlength{\abovedisplayskip}{3pt}
\setlength{\belowdisplayskip}{3pt}
\label{att_Eq}
\operatorname{Attention}(Q, K, V)=\operatorname{softmax}\left(\frac{Q K^{T}}{\sqrt{d}}\right) V
\end{equation}

The target user embedding $p_u$ is taken as $Q$, and the recipes representations $z_m$ are set as $K$ and $V$. The target meal representation $y_{m} \in \mathbb{R}^{1\times d}$ is defined as Eq(\ref{ym}):

\begin{equation}
\setlength{\abovedisplayskip}{3pt}
\setlength{\belowdisplayskip}{3pt}
\label{ym}
y_m = \operatorname{Attention}(p_u, z_m, z_m)=\operatorname{softmax}\left(\frac{p_u z_{m}^{T}}{\sqrt{d}}\right) z_m
\end{equation}

Since then, we have obtained the vector representation of the target meal $y_m$, which will be used later in meal prediction.

\subsection{Category-Wise User Representation}

In this subsection, our model learns category-wise user preferences by aggregating bottom-up from the recipe level to the category level and finally to the user level. 
This part is shown on the right of Figure \ref{framework}. 

Because we utilize user's historical interacted meals to estimate user preferences, we first process user input (omitted in Figure \ref{framework}). 
The number of user's interacted meals is not constant, so we set $l$ to be the maximum number of user's interacted meals the model can handle. If the number is greater than $l$, we randomly select $l$ meals. If the number of meals is less than $l$, a padding meal is repeatedly added to the user's meal list until the length is $l$. A meal contains three recipes for three categories, so the input of the user is a matrix $D_u \in \mathbb{N}^{|\mathcal{C}| \times l}$, where each element in $D$ is a recipe ID. Similar to meal recipe representation, we use recipe embedding matrix $E$ and category embedding matrix $O$ to get the representations of the user input $z_u$:

\begin{equation}
\begin{aligned}
\setlength{\abovedisplayskip}{3pt}
\setlength{\belowdisplayskip}{3pt}
\label{user_input}
z_u = 
\begin{bmatrix}
 z_{u-1_{a}} & z_{u-2_{a}} & \cdots & z_{u-l_{a}}\\
 z_{u-1_{s}} & z_{u-2_{s}} & \cdots & z_{u-l_{s}}\\
 z_{u-1_{d}} & z_{u-2_{d}} & \cdots & z_{u-l_{d}}
\end{bmatrix} \in \mathbb{R}^{|\mathcal{C}|\times l \times d}
\end{aligned}
\end{equation}

$z_{u-i_t}$ is the representation of the recipe with category $t$ in the $i$-th meal in the user's meal list, where $i \in [1,2,\dots, l]$ and $t \in  \mathcal{C}$.
Note that the $i$-th column of the matrix is the $i$-th meal in the user's interacted meal list, and the three rows of the matrix correspond to the recipes of the three categories.

\subsubsection{Recipe-Level Aggregation for Category}
We can get a summarized category representation (the user's preference for a category) by aggregating all the recipes that the user likes in the category. 
In order to reduce the irrelevant information and focus on the useful information about the target meal, we use the category representations of the target to guide the recipe-level aggregations for categories. The process involves extracting by category at the recipe level. The category representation of the user in category $t$ is defined in Eq(\ref{cut}):

\begin{equation}
\setlength{\abovedisplayskip}{3pt}
\setlength{\belowdisplayskip}{3pt}
\label{cut}
c_{u_{t}} = \operatorname{Attention}(z_{m_t}, z_{u_t}, z_{u_t})=\operatorname{softmax}\left(\frac{z_{m_t} z_{u_t}^{T}}{\sqrt{d}}\right) z_{u_t}
\end{equation}

where $c_{u_t} \in \mathbb{R}^{1\times d}$ and $t \in \mathcal{C}$. $z_{m_t}$ is the representation of the category $t$ of the target meal, that is, the recipe representation in the corresponding category. $z_{u_t}$ is a row for category $t$ in $z_u$, that is, the recipe representations of all the recipes in the category that the user likes.

After all categories are aggregated, we obtain the category representations of the target user $c_u$:

\begin{equation}
\setlength{\abovedisplayskip}{3pt}
\setlength{\belowdisplayskip}{3pt}
\label{zm}
c_u = 
\begin{bmatrix}
 c_{u_{a}}\\
 c_{u_{s}}\\
 c_{u_{d}}
\end{bmatrix} \in \mathbb{R}^{|\mathcal{C}|\times d}
\end{equation}

\subsubsection{Category-Level Aggregation for User}

Our model aggregates the category representations of the target user to obtain the category-wise user representation. Because different users place different emphasis on different categories, we can use the scaled dot-product attention for aggregation, where the user embedding $p_u$ is set as Q and category representations $z_m$ are taken as K and V. The category-wise user representation $x_u \in \mathbb{R}^{1\times d}$ is defined as Eq(\ref{xu}).

\begin{equation}
\setlength{\abovedisplayskip}{3pt}
\setlength{\belowdisplayskip}{3pt}
\label{xu}
x_{u} = \operatorname{Attention}(p_u, z_m, z_m)=\operatorname{softmax}\left(\frac{p_u z_{m}^{T}}{\sqrt{d}}\right) z_{m}
\end{equation}

In this section, we have proposed a hierarchical attention model to learn category-wise user representation, $x_u$, which contains important category information and focuses on information that pertains most to the target meal. 

\subsection{Recommendation}

In this part, the predicted score $\hat{r}_{um}$ between the target user and the target meal is calculated by the category-wise user representation $x_u$ and the meal representation $y_m$. Following \cite{NCF,AGREE,AttList}, we concatenate $x_u$, $y_m$ and their element-wise product as the input of the interaction model: 

\begin{equation}
\setlength{\abovedisplayskip}{3pt}
\setlength{\belowdisplayskip}{3pt}
\label{h}
h_0 = \begin{bmatrix}
 x_u\\
 y_m\\
 x_u\odot y_m
\end{bmatrix}
\end{equation}

After that, we apply a three-layer neural network to model user interactions and output the final prediction $\hat{r}_{um}$. Then, we can use this score to rank candidate meals to provide top-k recommendations.

\begin{equation}
\setlength{\abovedisplayskip}{3pt}
\setlength{\belowdisplayskip}{3pt}
\label{rum}
\left\{\begin{matrix}
 h_1=LeakyRuLU(W_1 h_0 + b_1)\\
 h_2=LeakyRuLU(W_2 h_1 + b_2)\\
 \hat{r}_{um}=LeakyRuLU(W_3 h_2 + b_3)
\end{matrix}\right.
\end{equation}

The Bayesian Personalized Ranking (BPR) loss \cite{bpr} is adopted to train our model.

\begin{equation}
\setlength{\abovedisplayskip}{3pt}
\setlength{\belowdisplayskip}{3pt}
\label{bpr}
Loss=\sum_{i=1}^{N} \sum_{s \in \mathcal{H}_{i}} \sum_{t \notin \mathcal{H}_{i}}-\ln \sigma\left(\hat{r}_{u_{i} m_{s}}-\hat{r}_{u_{i} m_{t}}\right)+\lambda\left\|\Theta\right\|_{2}^{2}
\end{equation}

where $\mathcal{H}_{i}$ denotes all meals interacted by user $u_i$. We can treat $(i, s)$ as a positive example and $(i, t)$ as negative, such that minimizing the loss function forces the prediction $ \hat{r}_{u_{i} m_{s}}$ to be larger than  $ \hat{r}_{u_{i} m_{t}}$. $\sigma(\cdot )$ is the sigmoid function, and $\Theta$ is the model parameter set of meal prediction task. $L_2$ regularization is applied to prevent over-fitting.

\begin{table*}[h]
\caption{Top-K recommendation performance comparison of different models on MealRec.}
\small
  \label{tab:result}
\begin{tabular}{c|cccccccc}
\hline
\multirow{3}{*}{Modle} & \multicolumn{8}{c}{MealRec}                                                                                                                                        \\ \cline{2-9} 
                       & \multicolumn{4}{c|}{$d$=5}                                                                 & \multicolumn{4}{c}{$d$=10}                                             \\ \cline{2-9} 
                       & HR@5            & NDCG@5          & HR@10           & \multicolumn{1}{c|}{NDCG@10}         & HR@5            & NDCG@5          & HR@10           & NDCG@10         \\ \hline
CFR \cite{meal_5}                   & 0.2921          & 0.1990          & 0.4171          & \multicolumn{1}{c|}{0.2394}          & 0.3714          & 0.2697          & 0.4717          & 0.3020          \\
DAM \cite{DAM}                   & 0.3321          & 0.2337          & 0.4527          & \multicolumn{1}{c|}{0.2725}          & 0.3835          & 0.2663          & 0.5162          & 0.3094          \\
AttList \cite{AttList}               & 0.4170          & 0.2981          & 0.5325          & \multicolumn{1}{c|}{0.3351}          & 0.4622          & 0.3473          & 0.5757          & 0.3811          \\
BGCN \cite{BGCN}                   & 0.3554          & 0.2421          & 0.4779          & \multicolumn{1}{c|}{0.2814}          & 0.4405          & 0.3119          & 0.5706          & 0.3540          \\
MIDGN \cite{MIDGN}                 & 0.4368          & 0.3108          & 0.5638          & \multicolumn{1}{c|}{0.3520}          & 0.4940          & 0.3680          & 0.6203          & 0.4191          \\ \hline
\textbf{CCMR}     & \textbf{0.4838} & \textbf{0.3537} & \textbf{0.6000} & \multicolumn{1}{c|}{\textbf{0.3913}} & \textbf{0.5524} & \textbf{0.4138} & \textbf{0.6927} & \textbf{0.4593} \\ \hline
\end{tabular}
\end{table*}

\section{Experiments}\label{sec-exp}
\subsection{Experiment Design}
\subsubsection{Evaluation Metrics}
The leave-one-out evaluation has been widely used in previous studies \cite{NCF,DAM,bpr,leave_one_out}. For each user, one of his or her interactions is randomly selected 
as the test set and remaining data is utilized for training. This evaluation method is also used in our experiments. Due to the time-consuming ranking of all items per user during evaluation, we follow a common strategy \cite{NCF,DAM} to randomly sample 99 meals that are not interacted with by the target user and rank the test meal among the 99 meals. To evaluate the top-K recommendation performance of models, two widely used metrics \cite{NCF,DAM,BGCN,MIDGN}, HR@K and NDCG@K, are employed to evaluate the recommendation quality of all models. HR (Hit Ratio) intuitively measures whether the test meal is present on the top-K list, and NDCG (Normalized Discounted Cumulative Gain) accounts for the position of the hit by assigning higher scores to hits at top ranks. 

\subsubsection{Baselines}

\begin{itemize}
  \item \textbf{GFR} \cite{meal_5}, a general framework for food recommendation, where collaborative learning is adopted to learn both user latent vectors and food latent vectors. 
  \item \textbf{DAM} \cite{DAM}, a multi-task model for bundle recommendation. It uses the attention mechanism aggregates items to obtain the representation of bundle, and utilizes user-item and user-bundle information simultaneously by share parameters between two tasks.
  \item \textbf{AttList} \cite{AttList}, a hierarchical self-attentive recommendation model leverages the hierarchical structure of items, bundles, and users to learn bundle-wise user preferences. 
  \item \textbf{BGCN} \cite{BGCN}, a graph-based model for bundle recommendation that reconstructs the user-item and user-bundle interactions and item-bundle affiliation into the graph. This model utilizes the graph neural network’s powerful ability to learn the user and bundle representations from complex structures. 
  \item \textbf{MIDGN} \cite{MIDGN} \footnote{available at https://github.com/CCIIPLab/MIDGN}, a model named Multi-view Intent Disentangle Graph Networks which extends DGCF \cite{DGCF} to the bundle recommendation task. It compares the intents learned from global and local views to represent the user’s preference. 
\end{itemize}
We have reproduced GFR based on their reported framework design and experimental settings. And other four baselines are implemented using the open source codes published by those authors and the parameters in the experiments follow the original settings in their works.

\subsubsection{Hyper-Parameter Settings}
Following all baselines, we implement our model through PyTorch and train it with Adam optimizer. The size of mini-batch is fixed at 1024, which is commonly used in baselines .For models with aggregate operations, the maximum number of user’s interacted meals $l$ is set as 64. Based on embedding size equal to 5 or 10, we conducted two experiments to compare their performance at different embedding sizes.

\subsection{Experimental Result and Analysis}

\subsubsection{Overall Performance}
The Top-K performance comparison of the proposed CCMR model and other state-of-the-art models under different embedding sizes are reported in Table \ref{tab:result}. As shown in the table, the proposed CCMR model achieved the best performance, followed by MIDGN and then AttList, BGCN, and DAM in order of dropping performance. The CFR model was sitting on the bottom with the poorest performance achieved in the experiments. 

GFR treats a meal as a single item without considering its recipe composition, which leds to the unsatisfied performance with 29.21\% in HR@5($d$=5). DAM is 4\% higher than CFB in HR@5 ($d$=5) mainly because it uses aggregated recipe representation to obtain the meal representation. However, DAM largely replied on the user ID to represent the user, ignoring the user preferences contained in the user's historical interaction. 

AttList and BGCN consider to aggregate the interacted meals of the target user through hierarchical neural networks and graph convolution networks, respectively, which is expected to estimate his or her preferences. AttList improves the metrics HR@5 and NDCG@5 in $d$=5 by 8\% and 6.44\% on the basis of DAM, but BGCN is only 2.23\% and 2.52\%.  
The main reason why AttList performs better than BGCN is because AttList uses the attention mechanism in aggregation to take into account the principle that different recipes and meals have different importance to users. BGCN uses the simple summation to aggregate recipes can only obtain non-personalized meal representations. 
MIDGN, as the current state-of-the-art model, reveals user-meal interactions from the perspective of intent, and it achieves good results. 
However, when AttList and MIDGN are applied in the meal recommendation, they focus on meal-wise user preferences and fail to introduce important category information, which results in its performance not being good enough.

Suggested by the experimental results measured by various evaluation metrics, our model CCMR outperforms all baselines. Especially at HR@10 and NDCG@10 at $d$=10, the proposed CCMR is 7.27\% and 4.02\% higher than the second best, MIDGN, which reflects the significant improvement achieved by our model over others in terms of recommendation accuracy and quality. Compared with other baselines such as CFR and DAM, CCMR notices the reflection of user's preferences on the user's historical interaction, thus improves the model's ability to realise user preferences, leading to a higher level performance. 

When we look at user historical interactions from the perspective of categories, we find another hierarchical structure of recipes, categories, and meals, which can be used to aggregate user historical interaction information. We use category as the intermediate node of aggregation. When aggregating recipes as categories, we consider that different categories of recipes have different unique characteristics. When category preferences are aggregated into user preferences, we can also take into account that users place different levels of importance on different categories. Compared to AttList, BGCN and MIDGN, the user's preference information for categories additionally modeled by CCMR also contributes a part to its superior performance to others in terms of recommendation. 

Before making recommendations, CCMR not only provides users with a personalized representation of the target meal, but also a specific, optimized user preference representation for the target meal. This is achieved by filtering information unrelated to the target meal. Bidirectional optimization of the recommendation process reduces the difficulty of final predictions. These features also contribute to the superior performance of CCMR to baseline models in bundle recommendation making, as demonstrated in the experimental results.


\begin{table}[!h]
\caption{Ablation Study (d=5).}
\small
  \label{tab:ablation}
\begin{tabular}{c|cccc}
\hline
Model & HR@5   & NDCG@5 & HR@10  & NDCG@10 \\ \hline
MW    & 0.4170 & 0.2981 & 0.5325 & 0.3351  \\
CW    & 0.4260 & 0.3063 & 0.5537 & 0.3471  \\
MW-F   & 0.4540 & 0.3290 & 0.5797 & 0.3696  \\
CCMR   & 0.4838 & 0.3537 & 0.6000 & 0.3913  \\ \hline
\end{tabular}
\end{table}

\subsubsection{Ablation Study}
The ablation study is designed based on the three observations in Section 4.2: \textit{meal-wise user preferences, category-wise user preferences}, and \textit{different recipes having different weights connecting to user preference for the target meal}. With the response to all three observations by our CCMR, we also developed three sub-models, CW, MW and MW-F. MW is a model like AttList~\cite{AttList} taking into account only the first observation, \textit{meal-wise user preferences}. CW moves a bit further and considers the first two observations. Alternatively, MW-F is designed with consideration of the first and third observations and performs additional information filtering on the meal-wise based MW. Comparing the performance of these sub-models with the CCMR model will gain us insight to the model for how the promising performance is achieved.


Ablation study are conducted with an embedding size of 5, and its results are shown in Table \ref{tab:ablation}. Compared with CW, MW has improved in all metrics, which indicates that category information is necessary for meal recommendation. Compared with CW, the performance of CCMR has a huge improvement, which also appears in the comparison of the results of MW and MW-F. CCMR and MW-F are obtained by adding filtering to user preferences based on the combination of CW and CW, respectively. The comparison of the experimental results between them confirms our third observation about filtering irrelevant information. In addition, the improvement of CCMR to CW is 5.84\% in terms of HR@5, which is higher than that of MW-F to MW at 3.70\%. This not only verifies the effectiveness of removing irrelevant information, but also shows that filtering at the category level is better than filtering at the meal level. 

To sum up the ablation experiments, we have discussed the effects of different strategies and combinations of them on the model, and finally have a conclusion that CCMR is the best solution for this experiment.

\section{Conclusion}\label{sec-conclusion}
In this paper, we introduced a meal recommendation dataset MealRec to address the lack of public datasets with user-meal interactions in this field. 
MealRec is built on recipe data and user review record data crawled from Allrecipes.com. We proposed a meal construction method that considers both explicit and implicit features and establish user-meal interactions based on user recipe interaction records. MealRec contains 1,500+ users, 7,200+ recipes, 3,800+ meals, 151,100+ user-recipe interactions and 46,700+ user-meal interactions. Furthermore, we proposed a category-constrained meal recommendation model and conducted comparative experiments with several state-of-the-art bundle recommendation methods on MealRec. The results proved the superiority of our model and demonstrated that MealRec is a promising testbed for the meal recommendation task. 
The conducted experiments on MealRec has also formed a new paradigm for evaluating bundle recommendation methods, which will guide the related research in the future for their practice in evaluation of newly proposed methods and algorithms. The original data and model code for MealRec are available to the research community for free access, and have potential to facilitate future meal recommendation research. MealRec can also be applied to research including, but not limited to, cross-modal retrieval of recipe images, food recognition and multimodal meal recommendation techniques. This will become our future work.


\bibliographystyle{ACM-Reference-Format}
\balance
\bibliography{references}

\end{document}